\documentstyle[preprint,prl,aps]{revtex}
\begin{document}
\draft
\preprint{\vbox{
\hbox{IFP-466-UNC}
\hbox{TRI-PP-93-16}
\hbox{hep-ph/9304294}
\hbox{April 1993}
}}

\title{
SSC Phenomenology of the\\
331 Model of Flavor
}

\author{Paul H. Frampton, James T. Liu, B. Charles Rasco}
\address{
Institute of Field Physics, Department of Physics and Astronomy,\\
University of North Carolina, Chapel Hill, NC 27599--3255, USA
}
\author{Daniel Ng}
\address{
TRIUMF, 4004 Wesbrook Mall\\
Vancouver, B.C., V6T 2A3, Canada
}
\maketitle
\begin{abstract}
{

The 331 model offers an explanation of flavor by anomaly cancellation
between three families.  It predicts three exotic quarks, $Q=D$, $S$,
$T$, and five extra gauge bosons comprising an additional neutral $Z_2$
and four charged dileptonic gauge bosons $(Y^{--},Y^-)$, $(Y^{++},Y^+)$.
Production of $Q\overline{Q}$, $QY$, $YY$ and $Z_2$ at the SSC is
calculated, and signatures are discussed.

}
\end{abstract}

\narrowtext
\paragraph*{Introduction.}
The standard model (SM) of strong and electroweak interactions is
extremely successful.  All experimental data are consistent with the
minimal version of the SM, and so theoretical extensions of the SM must
be motivated not, alas, by experiment, but by attempting to understand
features that are accommodated in the SM but not explained by it.  Perhaps
the most profound such feature is the replication of quark-lepton families
where the lightest family $(u,d,e,\nu_e)$ is repeated twice more,
$(c,s,\mu,\nu_\mu)$, $(t,b,\tau,\nu_\tau)$, with the only difference
between the families lying in the particle masses.

For consistency of a gauge theory, chiral anomalies must
cancel \cite{adler,bell,bouchiat}, and it
is a remarkable fact that this cancellation occurs between quarks and
leptons in the SM within each family separately.  For one family, this can
be used with only a few simple assumptions to deduce the electric charges of
the fundamental fermions including electric charge quantization without
recourse to the assumption of any simple grand unifying group containing
electric charge as a generator \cite{geng}.  The presence of more
than one family, however, is accommodated rather than explained in the SM. In
the 331 model \cite{pleitez,frampton}, which is an extension of the SM, one
obtains a theory with three extended families. While each extended family
has a non-vanishing chiral anomaly, the three families taken together
do not. This then offers a possible first step in understanding the flavor
question.

\paragraph*{The Model.}
The 331 model has gauge group
${\rm SU}(3)_c \times {\rm SU}(3)_L \times {\rm U}(1)_X$
(hence the name).
There are five additional gauge bosons beyond the
SM; a neutral $Z'$ and four dileptons, $(Y^{--},Y^-)$ with lepton number
$L = +2$ and $(Y^{++},Y^+)$ with lepton number $L = -2$.
Here $L = L_e + L_\mu + L_\tau$ is the
total lepton number; the 331 model does not conserve the separate
family lepton numbers $L_i$ ($i=e,\mu,\tau$).  The new $Z'$ will mix with
the $Z$ of the SM to give mass eigenstates $Z_2$ and $Z_1$, but the
singly-charged dilepton will not mix with the $W^\pm$ in
the minimal 331 model where total lepton number $L$ is conserved.

In the 331 model the leptons are the same in number as in the SM; however
the usual doublet--singlet pattern per family is replaced by
one antitriplet of electroweak ${\rm SU}(3)_L$.  Namely
the leptons in each family are in a $({\bf1},{\bf3}^*)_0$ under $(3_c,3_L)_X$.
The quarks are in $({\bf3},{\bf3})_{-1/3}$ and
$({\bf3}^*,{\bf1})_{-2/3,1/3,4/3}$ for the first and second
families and in $({\bf3},{\bf3}^*)_{+2/3}$ and
$({\bf3}^*,{\bf1})_{-5/3,-2/3,+1/3}$ for the third family \cite{twomodels}.
The result is that there are three new quarks in
the 331 model called $D$, $S$, and $T$ respectively with electric charges
$-4/3$, $-4/3$ and $+5/3$, and these provide targets of discovery at the
SSC.  From the manner in which the dileptons couple to
the heavy quarks, we see that the latter also carry lepton number. $D$ and $S$
have $L = +2$ and $T$ has $L = -2$.

The minimum Higgs structure necessary \cite{frampton} for symmetry breaking and
giving quarks and leptons acceptable masses is three complex ${\rm SU}(3)_L$
triplets, $\Phi({\bf1},{\bf3})_1$, $\phi({\bf1},{\bf3})_0$, and
$\phi'({\bf1},{\bf3})_{-1}$ and a complex sextet $H({\bf1},{\bf6})_0$.
The breaking of 331 to the SM is accomplished by a vacuum expectation
value (VEV) of the neutral component of the $\Phi({\bf1},{\bf3})_1$,
and this gives the only tree-level contribution to the masses of the
heavy quarks $D$, $S$, and $T$ and the principal such contribution to the
masses of the new gauge bosons $Z'$ and $Y$.  After ${\rm SU}(3)_L$ symmetry
breaking we are left with a three-Higgs doublet SM with new heavy
scalars, quarks
$D$, $S$, and $T$ and gauge bosons $Z'$ and $(Y^{\pm\pm},Y^\pm)$.  In this
letter, we examine the production and signatures of these new quarks and gauge
bosons.

\paragraph*{Gauge Boson Masses and Mixing.}
For triplets, we define the ${\rm SU}(3)_L\times {\rm U}(1)$ couplings $g$
and $g_X$
according to the covariant derivative
\begin{equation}
D_\mu=\partial_\mu-i g T^a W_\mu^a -i g_X XT^9 X_\mu\ ,
\end{equation}
where $T^a=\lambda^a/2$ and $T^9=\rm diag(1,1,1)/\sqrt{6}$ are normalized
according to ${\rm Tr}\, T^a T^b = {1\over2}\delta^{ab}$ and $\lambda^a$ are
the usual Gell-Mann matrices.  Note that $T^9=1/\sqrt{6}$
for ${\rm SU(3)}_L$ singlets.  $X$ is the ${\rm U}(1)_X$ charge of the
representations given above and is related to the usual hypercharge by
$Y/2=\sqrt{3}T^8+\sqrt{6}XT^9$ where $Q=T^3+Y/2$.

When ${\rm SU}(3)_L$ is broken to ${\rm SU}(2)_L$ by the VEV
$\langle\Phi^a\rangle = \delta^{a3} U/\sqrt{2}$, the new gauge bosons get
masses at tree level, $M_Y^2 = (1/4)g^2U^2$ and
$M_{Z'}^2 = (1/6)(2g^2 + g_X^2)U^2$, giving
\begin{equation}
\label{eq:massrel}
{M_Y\over M_{Z'}}=\sqrt{3g^2\over4g^2+2g_X^2}\ ,
\end{equation}
which is an ${\rm SU}(3)_L$ generalization of the $\rho$-parameter.

Using the matching conditions at the ${\rm SU}(3)_L$ breaking scale
\cite{ng}, $g_X$
is given by
\begin{equation}
\label{eq:match}
{g_X^2\over g^2}={6\sin^2\theta_W(M_{Z'})\over1-4\sin^2\theta_W(M_{Z'})}\ ,
\end{equation}
which allows us to rewrite
\begin{equation}
\label{eq:mymzp}
{M_Y\over M_{Z'}}={\sqrt{3(1-4\sin^2\theta_W(M_{Z'}))}\over
2\cos\theta_W(M_{Z'})}\ ,
\end{equation}
which gives a relationship between $M_Y$ and $M_{Z'}$
of $M_Y\le 0.26M_{Z'}$ for $\sin^2\theta_W(M_Z)=0.233$. This relationship
is specific to the minimal Higgs structure where ${\rm SU}(3)_L$ breaking is
accomplished only by triplet Higgs.  Where possible we will
discuss cross-sections for extended ranges of masses to allow
for the possibility of a more general non-minimal 331 model, but will
use Eq.~(\ref{eq:mymzp}) when a definite relationship between $M_Y$ and
$M_{Z'}$ is required.

At the electroweak scale, Eq.~(\ref{eq:massrel}) picks up
small corrections due to ${\rm SU}(2)_L$ breaking.  This allows the dilepton
doublet to be split in mass and also gives rise to $Z$--$Z'$ mixing.
Here, $Z$ is the standard mixture of $W^3$ and $Y$; whereas $Z'$ is the
gauge eigenstate orthogonal to $Z$ and $\gamma$.  Because the quarks in the
third family are in a different ${\rm SU}(3)_L$ representation, the $Z'$
coupling differentiates among families and hence there are tree-level
flavor-changing neutral currents (FCNC)
in the left-handed sector involving light quarks coupled to $Z'$
\cite{pleitez,ng}.

The $Z$--$Z'$ mixing arises from the mass matrix (in the $\{Z,Z'\}$ basis)
\begin{equation}
\label{eq:mixmat}
{\cal M}^2=\pmatrix{M_Z^2&M_{ZZ'}^2\cr M_{ZZ'}^2&M_{Z'}^2}\ ,
\end{equation}
where $M_Z^2=M_W^2/\cos^2\theta_W$ and $M_{ZZ'}^2$ are proportional to
${\rm SU}(2)_L$ breaking VEVs only.  Diagonalizing the mass matrix gives the
mass eigenstates $Z_1$ and $Z_2$ which can be taken as mixtures,
$Z_1=Z\cos\phi-Z'\sin\phi$ and $Z_2=Z\sin\phi+Z'\cos\phi$.  The mixing
angle, $\phi$, is given by
\begin{equation}
\tan^2\phi={M_Z^2-M_{Z_1}^2\over M_{Z_2}^2-M_Z^2}\ ,
\end{equation}
where $M_{Z_1}=91.173$GeV and $M_{Z_2}$ are the physical mass eigenvalues.

Because the entries in Eq.~(\ref{eq:mixmat}) arise from the same Higgs VEVs,
$M_{Z_2}$ and $\phi$ are related.  The mixing angle is
constrained to lie between the solid lines shown in Fig.~\ref{figmix}.
Additional constraints on this mixing arise from analysis of weak neutral
current
and precision electroweak measurements \cite{ng,langacker,liu}.  Shown on
Fig.~\ref{figmix} is a lower bound on $M_{Z_2}$ from FCNC \cite{ng}.
Since this bound comes from first--third family mixing, it is sensitive to
the values of the CKM parameters as well as new mixing angles coming from
the new quarks.

The matching condition, Eq.~(\ref{eq:match}), has an interesting
consequence --- namely $\sin^2\theta_W<1/4$ at the 331 breaking scale
\cite{frampton}.
Since $\sin^2\theta_W$ runs towards larger values at higher energies, this
provides an upper bound on the new physics.  Using the running of
$\sin^2\theta_W(M_{Z_2})$ for a three-Higgs doublet SM and demanding that
$\alpha_X(\mu)$ not be too strong puts an upper limit on $M_{Z_2}$.
For $\alpha_X(\mu)$ we impose, {\it faute de mieux}, an upper
bound $\alpha_X(\mu) < 2\pi$,
which implies $M_{Z_2} < 2200$GeV as shown in Fig.~\ref{figmix}.

An {\it indirect} lower limit on $M_{Z_2}$ follows from Eq.~(\ref{eq:mymzp})
and the empirical lower bound on the dilepton mass \cite{dng}, particularly
that coming from polarized muon decay \cite{carlson}.  Measurement of the
$e^+$ endpoint spectrum \cite{jodidio} gives the lower bound $M_Y>400$GeV
while the muon spin rotation technique \cite{beltrami} gives $M_Y>300$GeV
(both at 90\% C.L.).  Using the weakest lower bound, we find
$M_{Z_2} > 1400$GeV, which is shown in Fig.~\ref{figmix}.
In turn, this gives a strong limit on $Z$--$Z'$ mixing,
$-.02<\phi<+.001$, assuming the minimal Higgs sector.

The minimal 331 model is thus very predictive, and equally very easy to
rule out, giving a narrow window for the $Z_2$ mass between 1400GeV and
2200GeV and a corresponding window of between 300GeV and 430GeV for the
dilepton mass.

We note that the ${\rm U(1)}_X$ coupling constant $g_X$
diverges at a Landau pole less than one order of magnitude above
the $Z_2$ mass.
Because of this divergence, we assume
that $\alpha_X$ does not run above $M_{Z_2}$ (and hence $\alpha_X<2\pi$ is
always satisfied) when
evaluating the elementary cross-sections at SSC energies.
We expect such a behavior to be imposed by new physics above the $Z_2$ mass
in a more complete theory.

\paragraph*{$Z_2$ Production.}
Production of $Z_2$, which, due to the negligible mixing, is the same as $Z'$,
is dominantly by $q\overline{q}$ annihilation, and the
resulting cross-section is given by the solid line in Fig.~\ref{figZp}
(left-hand vertical axis).
The branching ratios for $Z_2$ decay are also shown
(right-hand vertical axis, dotted curves); here $M_Q=600$GeV has been
assumed.
The $\phi$ are the light physical scalars of the
three-Higgs SM, assumed to have a common mass
$M_\phi=200$GeV.
Note from Fig.~\ref{figZp} that the branching ratio into leptons is
extremely small ---
this is because leptons have $X=0$ and the $Z_2$ is dominantly in the
${\rm U}(1)_X$.
Nevertheless, the cross-section for $Z_2$ production is so
large that the charged lepton decay mode should be easily visible at the SSC.
The decay into a pair of dileptons will provide healthy
statistics for probing the $Z_2$.
Both these features are different from usual $Z'$ phenomenologies
\cite{robinett,leung,mahan,glashow,he}.
Forward-backward asymmetry would also be important to
distinguish this model from other $Z'$ models \cite{lanrobros}.

\paragraph*{Heavy Quark Production.}
Production of $Q\overline{Q}$ in $pp$
collisions proceeds through both the strong and the electroweak
interactions.  The strong interaction is through gluon fusion $gg\to
Q\overline{Q}$ and quark-antiquark annihilation
$q\overline{q}\to Q\overline{Q}$, and these
are identical to top quark $t\overline{t}$ production.
New production mechanisms in the 331 model, however, are available
in the electroweak sector by $t$-channel dilepton exchange
and, more importantly, by $s$-channel $Z_2$ exchange.  The latter mechanism
dominates production of $Q\overline{Q}$ for intermediate and large $M_Q$,
being an order of magnitude larger in cross-section than the strong
interactions for $M_Q\sim{1\over2}M_{Z_2}$ or above; this is because
$\alpha_X(\mu)$ is much larger than $\alpha_3(\mu)$ at these energies.

The decay modes of $Q$ are dictated by the conservation of lepton number.
Because, as already discussed, $D$ and $S$ have $L = +2$ they decay mainly by
$Q\rightarrow qY$. This leads in $pp\rightarrow Q\overline{Q}$ to two or
more jets from $q\overline{q}$ plus two lepton pairs from the $Y$s.
When both $Y$s are doubly-charged, the
leptons are like-sign, possibly resonant if $M_Q>M_Y$, and present a clear
signature.  For all $M_Q$ up
to 1TeV, the $Q$ production cross-sections are
in the multi-picobarn range and there are substantial event rates.

\paragraph*{Dilepton Production.}
By lepton number conservation, dileptons
must be produced either in pairs, $YY$, or in association with
a heavy quark $QY$.
Pair production occurs from $q\overline{q}$ annihilation through
$s$-channel $\gamma$, $Z_1$, $Z_2$ or $W$ exchange and $t$-channel heavy
quark exchange.
Depending
on the flavors in the $q\overline{q}$ initial state, one can produce
$Y^{++}Y^{--}$, $Y^{++}Y^-$, $Y^+Y^-$, or $Y^+Y^{--}$.  We have computed the
cross-sections using the leading log parametrization
of the structure functions given by Morfin and Tung \cite{morfin} evaluated at
$Q^2=\hat s$.  The results are depicted by the solid lines in
Fig.~\ref{figYYQY} for $M_Q=600$GeV.
The pairs $Y^{++}Y^{--}$ and $Y^+Y^-$ are produced similarly:
the cross-sections are nearly the same and are dominated by the $Z_2$
resonance below the threshold where $Z_2\to Q\overline{Q}$ becomes
kinematically allowed.

Associated production of a dilepton with a heavy quark is by
quark-gluon fusion, either by $s$-channel $q$ exchange or by $t$-channel
$Q$ exchange diagrams. The results for the production cross-sections
are depicted by the dotted lines in Fig.~\ref{figYYQY}.  These
are the cross-sections for production of a given dilepton in
association with an arbitrary heavy quark $Q=D$ or $S$.
In general, we find that associated production is
larger in cross-section than pair production, and depends more strongly on
the mass $M_Q$ of the heavy quark.

The decays of heavy quarks has already been discussed. The dilepton
decays either into a lepton pair or, if kinematically allowed, into
$qQ$ or scalar pairs. The doubly-charged dileptons $Y^{++}$ and $Y^{--}$
give the
especially simple signature of like-sign di-lepton pairs and is the most
striking effect predicted by the 331 model at the SSC. As discussed
above, the minimal version of the theory
demands that the dilepton mass is only just above the current
empirical lower bound 300GeV $< M_Y < 430$GeV so the SSC production
cross-sections are large, from one to several tens of picobarns.  Aside from
the $s$-channel production $e^-e^- \rightarrow Y^{--}$ \cite{paul},
the creation in a $pp$ collider is the best discovery
mode for dileptons.

Indirect searches for the existence of doubly-charged dileptons in
$e^+e^-$ colliders such as LEP-II and NLC have been previously
studied \cite{dng}.
Deviations from the SM expectations could show up even at LEP-II.
Direct searches have also been considered in $e^-p$
colliders \cite{agrawal}; although the cross-sections at HERA are
undetectably small, dileptons would be readily observable at LEP-II--LHC.

\paragraph*{Discussion.}
Because the dilepton mass is so restricted
by the minimal 331 model, its striking decay signature into like-sign
di-lepton pairs would be readily detectable at the SSC. Similarly
the $Z_2$, despite its higher mass, is easily within reach of the
SSC. These new gauge bosons would enrich the SM into
a new extended standard model which has the advantage of requiring
three families, rather than only accommodating them. Of course,
we shall again be at merely another stepping-stone to the final
theory.  But unless we confirm what the stepping-stone is, it is
not possible to guess with confidence what grand-unified theory
or even superstring theory it signals at extreme high energy
near or at the Planck scale. Once the families
are dealt with, however, there may be a better chance for the bold
extrapolation across an assumed desert to be successful.

Along these lines, the Landau pole in
$\alpha_X(\mu)$ mentioned above makes clear that new physics
beyond the minimal 331 model is crucial even before the desert
begins. What this new physics is should be ascertainable
or, at least, strongly restricted by the requirement of
embedding the theory in a simple or quasi-simple unifying
group and by the usual requirements about running couplings meeting
together at the unification scale. The anomaly cancellation in
the minimal 331 model already suggested a grand unified theory
as the origin, and now the Landau pole shows why failure to find
a simple unification for the minimal model was not surprising since
something extra is needed to make the abelian ${\rm U}(1)_X$ more
asymptotically free even before the long extrapolation towards the
Planck mass.  We expect that these considerations about even higher
energies will not affect the predictions of the minimal 331 model we have
presented at energies accessible to the SSC and hence await with interest
what the SSC will reveal.

We thank Ernest Ma for useful discussions.
This work was supported in part by the U.S. Department of Energy under
Grant No.~DE-FG-05-85ER-40219 and by the Natural Science and
Engineering Research Council of Canada.

\begin{figure}
\caption{Constraints on $M_{Z_2}$ and mixing angle $\phi$ from the Higgs
structure (solid lines).
The lower and upper bounds on $M_{Z_2}$ are explained in the text.
The cross-hatched region is allowed.}\label{figmix}
\end{figure}
\begin{figure}
\caption{$Z_2$ branching ratios (dotted lines) and production cross-section at
$\protect\sqrt{s}=40$TeV (solid line).  We have taken $M_Q=600$GeV and light
scalar masses to be 200GeV.}\label{figZp}
\end{figure}
\begin{figure}
\caption{Cross-sections for dilepton pair production (solid lines) and
associated production (dotted lines) at $\protect\sqrt{s}=40$TeV
for heavy quarks with mass $M_Q=600$GeV.}
\label{figYYQY}
\end{figure}


\begin{references}
\bibitem{twomodels} We use the choice of representations for the first
and third families made in
Ref.~\cite{frampton} as seems necessary for consistency with
flavor-changing neutral current constraints \cite{ng}.
\end{references}
\end{document}